\documentstyle[pra,aps,preprint,tighten]{revtex}

\begin{document}

\title{Universal Behaviour of Metal-Insulator Transitions in the  
p-SiGe System}

\author{ P.T. Coleridge, P. Zawadzki, A.S. Sachrajda,
    R.L. Williams and Y. Feng  }

\address{Institute for Microstructural Sciences, National Research
Council
of Canada,\\Ottawa, Ontario, K1A 0R6, Canada}


\maketitle

\bigskip
 {\footnotesize
Magnetoresistance measurements are presented for a strained p-SiGe quantum well
sample where the density is varied through the B=0 metal-insulator transition.
The close relationship between this transition, the high field Hall insulator
transition and the filling factor $\nu$=3/2 insulating state is demonstrated.
}

\bigskip
  
     The strained p-type SiGe system exhibits, in addition to the
normal integer quantum Hall effect (IQHE) transitions, an
insulating phase near filling factor $\nu$ = 3/2 \cite{dunford,ssc} and
a B=0 metal-insulator transition of the kind observed in high
mobility Si-MOSFETs \cite{kravchenko}. Results are presented here
that show the close relationship between these various transitions.

     Samples were grown by a UHV chemical vapour deposition
process.  An intrinsic Si layer is followed by a 40nm  Si$_{.88}$
Ge$_{.12}$ quantum well, a spacer layer of variable thickness and
a boron doped silicon layer. The quantum well is sufficiently
narrow that the lattice constant difference between the alloy and
the pure Si is all taken up by strain which means the heavy hole
band, characterised by a  $|M_J|$ = 3/2 symmetry, is well separated
from other bands. The g-factor is large (of order 6) and depends
only on the perpendicular component of magnetic field \cite{ff,ptc1}.
  At high fields the spin splitting is further enhanced 
by exchange resulting in a fully polarised ``ferromagnetic'' 
spin system at $\nu$ = 2. The ratio of the transport 
and quantum lifetimes, determined from
the Shubnikov-de Haas oscillations, is about one \cite{PRB} showing
that the disorder is dominated by a short-ranged scattering
potential.  

       Transitions between the N$_L$ and N$_{L+1}$ integer quantum
Hall states are well described (in units of e$^2$/h) by
\cite{preprint1}
  
\begin{equation}  
       \sigma _{xx} =  s / (1 + s^{2}), \; \; \; \; \; \;
    \sigma _{xy} = N_L  + 1 / (1 + s^{2}).
\label{eq1}  
\end{equation}  
with a scattering parameter s, (which can be identified with the
Chern-Simons boson conductivity $\sigma_{xx}^{(b)}$ \cite{KLZ}) 
given by

\begin{equation}  
             s     =  \exp [ (\nu_c -\nu) ( T_0 / T)^{\kappa} ].  
\label{eq2}  
\end{equation}
Here $\nu_c$ is the critical filling factor and the exponent
$\kappa$ is close to 3/7 at low temperatures. For N$_L$ =0, ie at
the termination of the Quantum Hall sequence, this gives

\begin{equation}
              \rho_{xx}  = s  , \; \; \; \; \; \; \rho_{xy} =  1
\label{eq3}
\end{equation}
corresponding to a quantised Hall insulator.

       Near $\nu$ = 1.5 another insulating phase is observed in
many samples. The presence of this phase depends on density,
disorder and on magnetic field tilt \cite{dunford,ssc}. An
activation analysis \cite{ssc} correlates insulating behaviour with
the existence of degenerate spin states at the Fermi level: that is
it is suppressed when the ferromagnetic polarisation of the spins
at $\nu$ = 2 persists through $\nu$ = 1.5 as the field is
increased. This can be demonstrated to be a re-entrant transition,
growing out of the $\nu$ = 1 IQHE state although, but when it is
very strongly developed it appears to grow directly out of 
the $\nu$ = 2 or 3 states. 

     Figure 1a shows the $\nu$=3/2 and high field Hall insulating
transitions measured using a two terminal technique in a sample
with a density  of 1.4 $\times$10$^{11}$cm$^{-2}$. A scaling plot
of this data (fig.1b) shows $\rho_{xx}$ plotted against $ (\nu_c -
\nu) / T^{\kappa} $ with $ \kappa $ = 3/7. There is good agreement
with eqn.2 in both cases with slightly different values of T$_0$.
The scaling deteriorates in the $\nu$ = 3/2 insulating phase
because of the close proximity of the two critical points.
       
     For the B=0 metal-insulator transition the temperature
dependence of $\rho_{xx}$ in the insulating phase is well
described over several orders of magnitude, by $\rho_c \exp[
(T_0/T)^{n}]$ with $\rho_c \sim $ 0.5h/e$^2$ and n $\sim $ 0.4
\cite{PRB}. In the metallic phase, at low T, it is of the general
form 

\begin{equation}  
      \rho_{xx}(T)     = \rho_0  +  \rho_1 \exp [ - (T_1/T)^{p} ].
\label{eq4}  
\end{equation}  
As is commonly observed in these systems \cite{zzz,Hanein}, for 
densities near the critical value the resistance does not always 
increase monotonically and often has a maximum and 
a tilted separatrix between the metallic and 
insulating phases. This makes it difficult
to independently determine the prefactor $\rho_1$ and the exponent
p by fitting to eqn.4. Choices of p=1 (and  $\rho_1$ small) or
alternatively p $\approx$ 0.4 (with $\rho_1$ then of order 0.5
h/e$^2$) are equally successful. In each case, however, the parameter
T$_1$ varies with density and there is a general similarity to
eqn. 2.

       Although the temperature dependence in the ``metallic''
phase is dominated by activated behaviour there is also evidence of
weak localisation \cite{wklocn}. Within experimental error, a ln(T) 
term cannot be detected directly, but the negative
magnetoresistance around B = 0 and positive behaviour at higher
fields, can be consistently interpreted in terms of a sum of weak
localisation and Zeeman contributions. The value of F$^{\ast}$
\cite{L&R} extracted in this way is large (2.45). Combined with
the cancellation between the two terms it leads to a coefficient for
the ln(T) behaviour which is close to zero, consistent with the
experimental data. 

       Figure 2 shows magnetoresistance data 
for a sample where the density was
varied, by exploiting a persistent photoconductivity effect,
through the B=0 critical value. The zero field resistivities
(figure 3) show the critical density is 7.8 $\times$
10$^{10}$cm$^{-2}$ . For the highest density trace (figure 2a) there
 are three fixed points corresponding respectively to transitions into 
the $\nu$ = 3/2 insulating phase, into the  $\nu$ = 1 QHE state and
into the high field Hall insulating phase. As the density is
reduced the first transition disappears (or moves to a much lower
field); this is followed, in the next trace by the simultaneous 
disappearance of the two higher field fixed points so, at 
the lowest density, the temperature dependence over the 
whole field range is insulating. The low
field Hall resistance is well defined through the whole range of 
densities. This indicates a Hall insulator with $\sigma_{xx}$ and
$\sigma_{xy}$ both diverging but with the 
ratio $\rho_{xy} = \sigma_{xy} / \sigma_{xx}^{2}$ 
taking the classical value (B/pe, where p is the
density).  At higher fields it retains this classical behaviour 
until it becomes quantised near $\nu$=1 and approximate quantisation
then continues well into the high field insulating state, consistent with 
eqn.3.

       The critical resistivities for all three transitions: the B=0
transition, the $\nu$ = 3/2 transition, and the high field Hall
insulator transition are approximately 0.5 h/e$^2$. In contrast to
the situation in p-GaAs[\cite{Hanein}] the resistivity at the high
field transition point (and also for the $\nu$ = 1.5 transition) is
almost independent of density. Furthermore, the B = 0 transition 
is unchanged by magnetic field. Again, this is in contrast to the
situation in p-GaAs where the B=0 transition transforms smoothing 
into the IQH effect transition. This difference in behaviour is
probably a consequence of the strong spin-coupling in p-SiGe 
which quenches the independent degree of freedom of the spins.

       The behaviour is summarised in a phase diagram (figure
4) which is to some extent schematic. At
high densities a well defined Landau level structure is observed 
with the re-entrant $\nu$ = 3/2 insulating phase. At lower
densities this is washed out in a region where $\Gamma$ 
(the Landau level broadening) is larger than 
$\hbar \omega_c$ (the Landau level spacing), but the high field 
and $\nu$ = 3/2 insulating phases
persist. At the lowest density the behaviour is insulating, over
the whole field range, with no clear distinction between the 
three types of insulating behaviour.  The well-known ``floating-up'' 
of the lowest Landau level is shown. In this case the 
condition $\Gamma$ = $\hbar \omega_c$ is the same as the 
criterion for the appearance of the insulating behaviour
 k$_F${\it l} = 1 (where {\it l} is the mean free path). For higher Landau 
levels, however, this is not the case and the disappearance 
of the Landau levels must be associated more with the dominance 
of the disorder than with ``floating-up''.

     In all cases spin plays an important role. For the B = 0 
transition, in the insulating phase, this is demonstrated by the 
insensitivity to magnetic field; in the metallic regime it is
 presumeably is the cause of the large value of F$^{\ast}$. The 
high field quantised Hall insulating state is, by definition, 
spin polarised and spins also seem to play a role in the
formation of the $\nu$ = 3/2 insulating state.

\newpage

\begin{figure} [p]  
\vspace*{5.5cm}  
\includegraphics{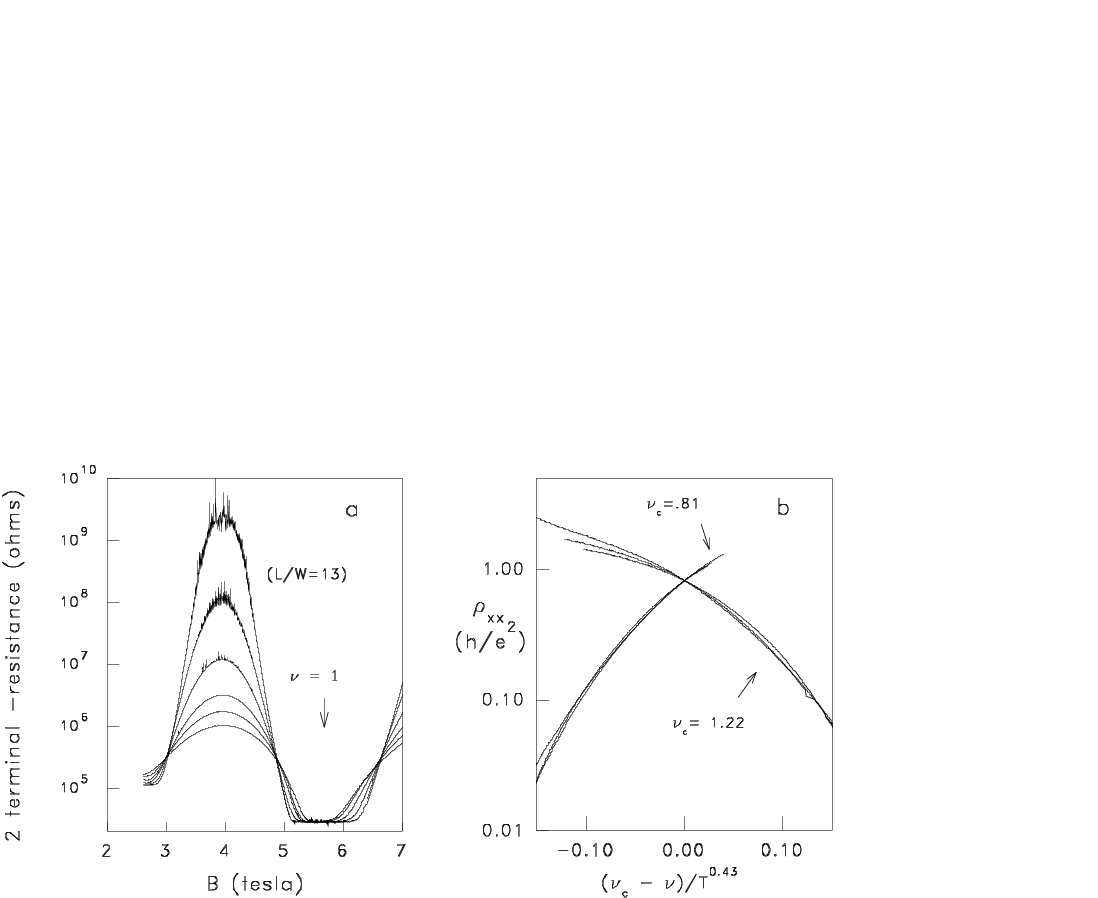}  
\caption  
{ (a) Two terminal resistance showing the $\nu=3/2$ insulating 
phase with a critical point at $\nu_c$=1.22 and the high field 
Hall insulating transition (with $\nu_c$=.81) at T = 75, 120, 220,
400, 600 and 900mK. (b) Scaling plot for the 120,220 and 400mK 
data from (a).
 }  
\label{fig1}  
\end{figure} 
  
\begin{figure} [p]  
\vspace{13cm}  
\includegraphics{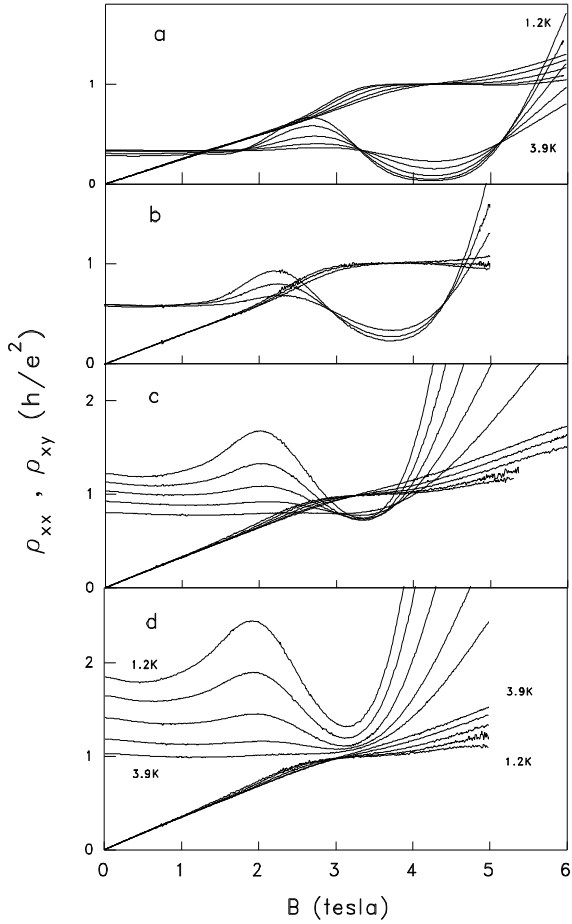}  
\caption  
{ Magnetoresistance at temperatures of 1.2, 1.55, 2.15, 3.0 and 
3.9K for densities (in units of 10$^{10}$cm$^{-2}$) of:(a) 9.7,
(b) 8.1 (1.2 - 2.15K data only) , (c) 7.0, (d) 6.6. 
}  
\label{fig2}  
\end{figure}  
  
\newpage

\begin{figure} [p]  
\vspace*{6.cm}  
\includegraphics{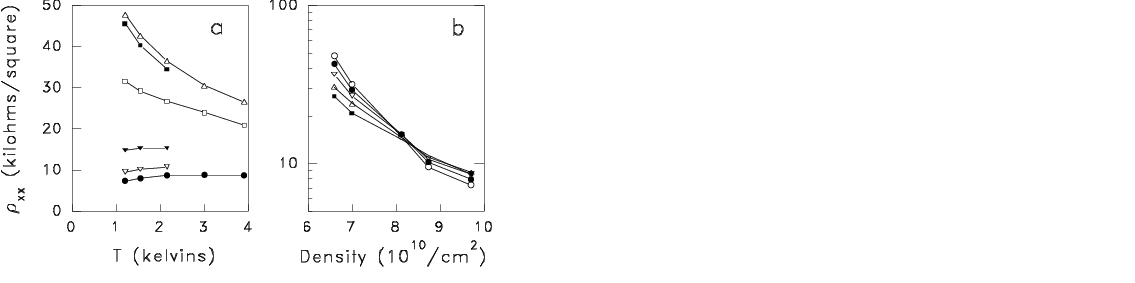}  
\caption  
{Zero field resistivity (a) as a function of T; (b) as a function
of density.
}
\label{fig3}  
\end{figure} 

  
\begin{figure} [p] 
\vspace{8cm} 
\includegraphics{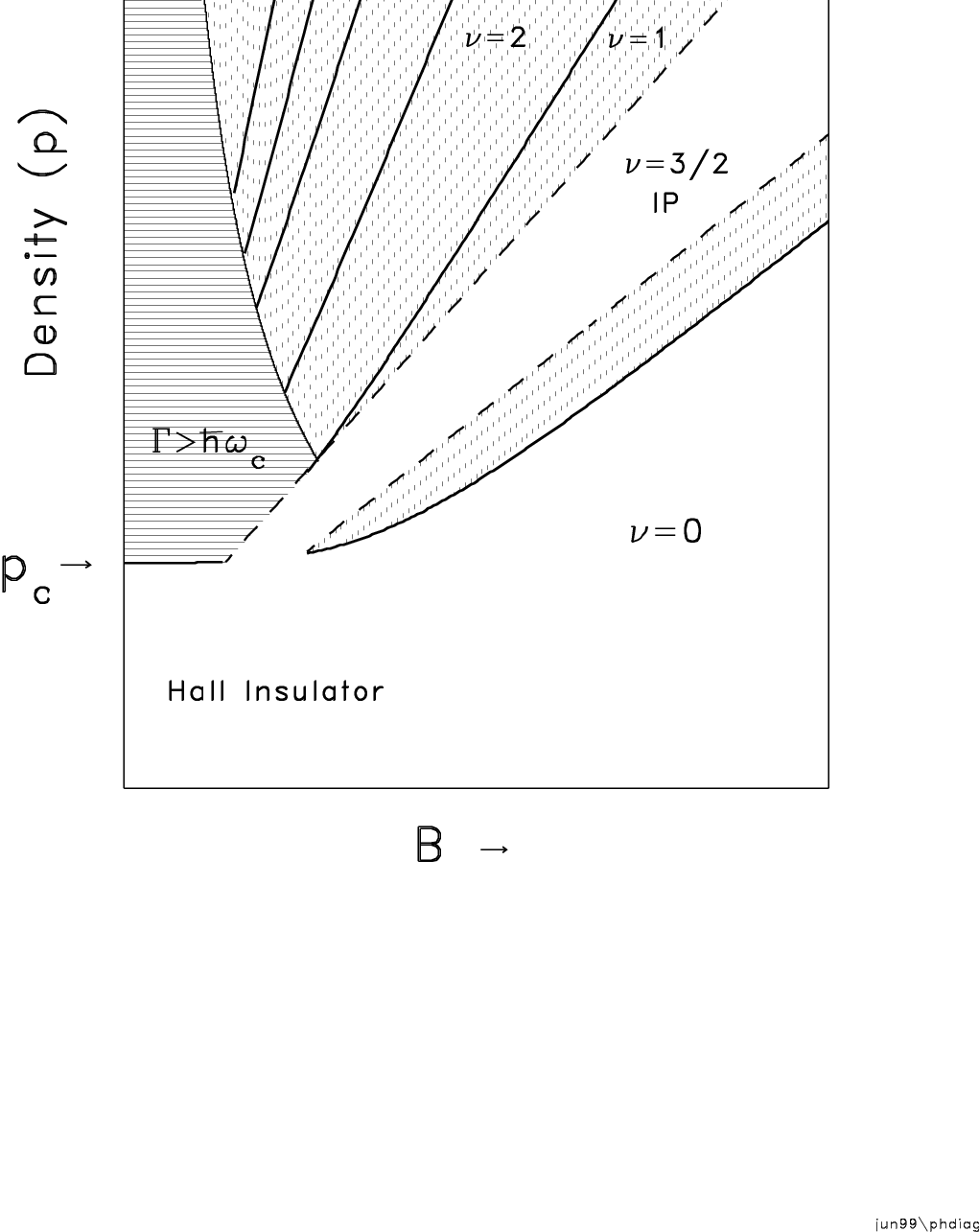}
\caption
{Schematic phase diagram showing Landau levels, the re-entrant 
insulating phase at $\nu$=3/2 and the high field insulating phase.  
p$_c$ is the B=0 critical density. Note also the region 
where the Landau level broadening is larger than the 
spacing.  
}  
\label{fig4}  
\end{figure}

\end{document}